\documentclass[useAMS,usenatbib]{mn2e}
\usepackage{graphicx}
\usepackage{amssymb}
\usepackage{amsmath}
\usepackage{algpseudocode}
\usepackage{multicol}
\usepackage{longtable}
\usepackage{algorithm}
\usepackage{url}
  
\title[Reduction of astrochemical networks]{Chemical complexity in astrophysical simulations: optimization and reduction techniques}

\author[T. Grassi et al.]{T. Grassi$^{1}$\thanks{Corresponding author: tommasograssi@gmail.com}, S. Bovino$^{2}$, D. Schleicher$^{2}$, F. A. Gianturco$^{1}$\\
$^{1}$Department of Chemistry, Sapienza University of Rome, P.le A. Moro, 5, 00185 Roma\\
$^{2}$Institut f\"ur Astrophysik Georg-August-Universit\"at, Friederich-Hund Platz 1, 37077 G\"ottingen\\}

\begin{document}
\newcommand{\ith}{$i$th }
\newcommand{\jth}{$j$th }
\newcommand{\nth}{$n$th }

\newcommand{\dd}{\mathrm d}
\newcommand{\mA}{\mathrm A}
\newcommand{\mB}{\mathrm B}
\newcommand{\mC}{\mathrm C}
\newcommand{\mD}{\mathrm D}
\newcommand{\mE}{\mathrm E}
\newcommand{\mH}{\mathrm H}
\newcommand{\mSi}{\mathrm Si}
\newcommand{\mO}{\mathrm O}
\newcommand{\cmc}{\mathrm{cm}^{-3}}
\newcommand{\real}{\mathbb R}
\newcommand{\superscript}[1]{\ensuremath{^{\scriptscriptstyle\textrm{#1}\,}}}
\newcommand{\trader}{\superscript{\textregistered}}

\newcommand\mnras{MNRAS}
\newcommand\apj{ApJ}
\newcommand\apjs{ApJS}
\newcommand\aap{A\&A}
\newcommand\apss{Ap\&SS}
\newcommand\osu{osu\_01\_2007}

\date{Accepted *****. Received *****; in original form ******}

\pagerange{\pageref{firstpage}--\pageref{lastpage}} \pubyear{2012}

\maketitle

\label{firstpage}

\begin{abstract}
Chemistry plays a key role in the evolution of the interstellar medium (ISM), so it is highly important to follow its evolution
in numerical simulations. However, it could easily dominate the computational cost when applied
to large systems. In this paper we discuss two approaches to reduce these costs:  (i) based on computational strategies, and 
(ii) based on the properties and on the topology of the chemical network. The first
methods are more robust, while the second are meant to be giving important information on the structure of large, complex networks.

\noindent We first discuss the numerical solvers for integrating the system of ordinary differential equations (ODE) 
associated with the chemical network, and then we propose a buffer method that decreases the computational time spent in solving the ODE system. 
We further discuss a flux-based method that allows one to determine and then cut \emph{on the fly} the less active reactions.

\noindent In addition we also present a topological approach for selecting the most probable species that will be active during the chemical evolution,
thus gaining information on the chemical network that otherwise would be difficult to retrieve. 
This topological technique can also be used as an \emph{a priori} reduction method for any size network.

\noindent We implemented these methods into a 1D Lagrangian hydrodynamical code to test their effects: 
both classes lead to large computational speed-ups, ranging from $\times2$ to $\times5$. 
We have also tested some hybrid approaches finding that coupling the flux method with a buffer strategy gives
the best trade-off between robustness and speed-up of calculations.  

\end{abstract}

\begin{keywords}
astrochemistry -- ISM: evolution, molecules -- methods: numerical.
\end{keywords}

\section{Introduction}
Chemistry plays a key role in many astrophysical environments (e.g. \citealt{GalliPalla98}, \citealt{Nelson1999}, \citealt{Omukai2005},  \citealt{MerlinChiosi07}, \citealt{Meijerink2005}, \citealt{Tesileanu2008}, \citealt{Gnedin2009}, \citealt{Glover2010}, \citealt{Yamasawa2011}, \citealt{Grassi2011}). 
It regulates the cooling and the fragmentation of the interstellar gas \citep{Jappsen2005} and it is one of the most powerful diagnostic way to relate with the observational signatures coming from the last generation of telescopes like ALMA\footnote{http://www.almaobservatory.com} and Herschel\footnote{http://herschel.esac.esa.int/}.
Unfortunately tracking its evolution needs large computational resources.
The complexity of the problem arises from its mathematical representation: a network of chemical reactions has its counterpart in a system of coupled ordinary differential equations (ODE) 
that is often stiff, {since it is a problem with two or more very different scales of the independent variable on which
the dependent variables are changing (e.g. see Sect. 16.6 of \citealt{NumericalRecipes}).} 
The solution of such a systems may becomes too expensive computationally, especially when the number of reactions is large, or when the system is formed by many elements of fluid.

Over the years several approaches have been proposed to solve this problem: reducing the number of reactions/species by making physical or chemical assumptions is one of the most used. 
The number and the kind of reactions being removed depends on the astrophysical environment that one wants to simulate (e.g. \citealt{Nelson1999}). 
This method can be useful when the network is small, but it is hard to apply when a larger network is employed,
leading to large errors with complex networks.

Another widely used approach is to reduce the number of non-equilibrium species (e.g. \citealt{Glover2010}) while treating 
the remaining species as being in equilibrium. This method decreases the chemical time-step and thus the computational cost, but it is problem-dependent
since the selection of the non-equilibrium species depends on the environment one wants to simulate.

In this paper we discuss a set of methods that can be used to increase the feasibility of accurate simulations based on large chemical networks
while also providing some key information on the structure of the network.

We introduce the problem in Sect.\ref{sect:problem_overview}, then we consider two classes of reduction strategies: 
(i) the first is based on methods more related to computer science and we call them ``pure'' computational 
strategies (see Sect.\ref{sect:pure_strategies}),  (ii) while the second set is based on the analysis of the properties 
and on the topology of the chemical network (Sect.\ref{sect:reduction_methods}).
In Sect.\ref{sect:test} we will show tests on some of the methods previously discussed.  
\section{Overview of the problem}\label{sect:problem_overview}
A chemical network is composed of a number of reactions each generally represented by $\mA + \mB \to\mC + \mD$, and the probability that a reaction occurs
is related to a rate coefficient $k_j(T)$ that depends on the gas temperature $T$. In this framework each species evolves following an ordinary differential equation (ODE)
that takes into account the reactions that form and destroy the \ith species
\begin{equation}\label{eqn:ODE}
	\frac{\dd n_i}{\dd t} = \sum_{j\in form} k_j(T) n_{r1(j)}n_{r2(j)} - n_i\sum_{j\in destr} k_j(T)n_{r1(j)} \,,
\end{equation} 
where $n_{r1(j)}$ and $n_{r2(j)}$ are the number densities of the first and the second reactants of the \jth reaction, and  the subscript $form$ and $destr$ represents 
the set of the reactions that form and destroy the \ith species. Note that for three-body reactions we must multiply the first part
 (and/or the second, depending on the reaction) of the RHS term by
the number density of the third reactants, i.e. $n_{r3(j)}$. The network is then described by a system of ODEs.

From a computational point of view the problem of solving a system of ODEs resides mainly in building the RHS term of Eqn.(\ref{eqn:ODE}), 
while the total number of equations has a smaller (but non-negligible) impact on the solver efficiency \citep{Tupper2002}. 
To achieve this we can either remove the species, decreasing the number of ODEs
and hence the dimensionality of the problem, and/or reduce the number of reactions, thus diminishing the 
time spent to build the RHS of Eqn.(\ref{eqn:ODE}). Both approaches will be discussed in Sect.\ref{sect:reduction_methods}.

\section{Pure computational strategies}\label{sect:pure_strategies}
In this Section we discuss the strategies that are less related to the properties and the features of the chemical network, but allow to improve 
the computational performance of solving a system of ODEs aimed at tracking the chemical evolution. We first present (i) an analysis 
of the different solvers,  then (ii) the buffering method, and finally (iii) a general discussion on standard computational techniques. 
\subsection{Numerical solvers}
Astrochemical networks are represented by a system of stiff ODEs, which requires appropriate solvers. 
Due to the numerical instability of the system, one of the most used solvers is the implicit backward differencing (BDF) scheme,
but depending on its implementation it can greatly affect the computational performance. 
A large number of solvers have been proposed over the years like Rosenbrock, 
Bulirsch-Stoer-Bader-Deuflhard, implicit Runge-Kutta, and Gears  \citep{NumericalRecipes}, 
but unfortunately these methods have poor performance  when applied to astrochemical networks 
since a large number of iterations and function evaluations are required to integrate the ODE system. 
A good way to solve this problem is to use the well-established ODEPACK/SUNDIALS package\footnote{http://computation.llnl.gov/casc/sundials/main.html} \citep{Hindmarsh2005}, a solver suite  based on
GEARS and LSODE that allows to solve stiff ODE systems more efficiently. The DVODE fits our class of problems as well, but nevertheless, 
when the system presents a sparse Jacobian the DLSODES solver \citep{Hindmarsh83}, which has the capability of handling sparse matrices, 
is more efficient.
DLSODES has been already discussed in \citet{Nejad2005}, and we
have tested here the two solvers on a larger network \citep{Wakelam2008} involving 452 species and more than 4000 reactions.
The DVODE and the DLSODES will be tested in Sect.\ref{sect:test_solver}.

\subsection{Buffering}\label{sect:buffering}
In the framework of hydrodynamical simulations one usually solves the chemical ODEs for each fluid element (i.e. each gas particle or grid cell), 
the number of which can easily exceed $10^6$, depending on the resolution: 
for this reason reducing the number of calls to the solver saves
a large amount of CPU-time. This allows us to take advantage of the fact that we can avoid to solve the chemical ODEs for fluid elements 
which have chemical conditions (e.g. temperature,
species numerical densities, \dots) similar to those that were already calculated, since they will show the same evolution with time. This comes from the fact that a 
system of ODEs is a Cauchy's problem that
depends on the set of initial conditions and on the differential equations which are weighted by the values of the reaction rates.
 
The buffering method we propose consists of 
storing the already calculated chemical results. 
If the initial conditions of a given fluid element matches the initial conditions of a buffered element we can avoid to call the solver again, and
use instead the stored evolution. In the opposite case we call the solver to evolve the fluid element and we store the initial and the final conditions in the buffer (see also Algorithm \ref{pseudo:buffer} discussed below). 
\begin{algorithm*}\caption{ - Pseudocode for the buffering method proposed here. Note that the term \emph{particle} is the same as \emph{fluid element}. 
	The aim of this psuedocode is to find the final conditions $\mathbf{\hat x}$ of a particle with initial conditions $\mathbf{x}$.
	See details in Sect.\ref{sect:buffering}.}
\label{pseudo:buffer}
\begin{multicols*}{2}
	\begin{algorithmic}[1]
		\State $\mathbf{x}\gets$ initial conditions
		\For {$(j=N,1)$}
				\State found$\gets$True
				\If{($|\dd t_j - \dd t|>\epsilon_t$)}
					\State found$\gets$False
					\State skip to next fluid element
				\EndIf
				\For {$(i=1,N_c)$}
					\If{($|x_i-x_{ji}|/x_i>\epsilon$)}
						\State found$\gets$False
						\State exit from initial condition loop
					\EndIf
				\EndFor
				\If{(found)} 
					\State $\mathbf{\hat x} \gets \mathbf{\hat x_j}$
					\State exit from buffer loop
				\EndIf
		\EndFor

		\If{(not found)}
			\State $\hat{\mathbf x}\gets$ solver$(\mathbf{x},\dd t)$
			\State $\hat{\mathbf x}_{N+1}=\hat{\mathbf x}$
			\State ${\mathbf x}_{N+1}={\mathbf x}$
			\State $\dd t_{N+1}=\dd t$
			\State $N=N+1$
		\EndIf
	\end{algorithmic}
\columnbreak

\begin{algorithmic}
	\State // $\mathbf x$ are the initial conditions of a given particle
	\State // loop over buffer particles ($j=N\to1$)
	\State // set default value of found flag
	\State // compare $\dd t$
	\State // set found flag to false
	\State // skip to the next buffer particle (next $j$)
	\State // 
	\State // loop over initial conditions ($i$)
	\State // check similarity for the \ith initial condition
	\State // set found flag to false
	\State // breaks loop over initial conditions ($i$)
	\State // 
	\State // end loop over initial conditions
	\State // if particles match
	\State // load final conditions from buffer
	\State // go to the next cell (break loop $j$)
	\State // 
	\State // end loop over buffer particles
	\State // if the particle is not in the buffer
	\State // compute final conditions with solver
	\State // store final conditions in the buffer
	\State // store inital conditions in the buffer
	\State // store time step in the buffer
	\State // increase the size of the buffer
	\State // 
\end{algorithmic}
\end{multicols*}
\end{algorithm*}

To determine if two particles are similar we loop over the initial conditions and we check if the expression 
$s_i=|x_{ij} - x_i|/x_i<\epsilon$ is true, where $x_{ij}$ 
is the \ith initial condition of the \jth particle of the buffer, and $x_i$ is the \ith initial condition of the test particle. 
It is worth noting that we also need a constraint on the time-steps of the two particles, namely $|\dd t - \dd t_j|<\epsilon_t$, where
$\epsilon_t$ is the time threshold, and
in the tests of Sect.\ref{subsect:test_buffer} we use $\epsilon_t\approx0.1$ yr. The time threshold is chosen in order to have a good
approximation over the time-step and for this reason its value must be a fraction of the mean hydrodynamical time-step, depending
on the desired accuracy. 
From our test we found that a small $\epsilon_t$ provides accurate results even with larger
values of $\epsilon$.

The total error is given by the chosen $\epsilon$ and $\epsilon_t$, and it also depends on the chaotic 
behaviour of the chemical network. If the system is unstable to small perturbations of the initial parameters 
we must choose a very small $\epsilon$ thereby reducing the advantage of using the buffer. 

In order to increase the speed-up of the method we scan the buffer from the most recent values to those stored at earlier stages, 
since it is more likely that the last particles stored are similar to the test
particle. This increases the probability of finding a similar particle earlier during the scan,
allowing us to break the cycle after a few comparisons: this increases the advantage of using this method.

Since the size $N$ of the buffer increases while the simulation is running, we must define a maximum buffer size, namely $N_\mathrm{max}$,
that is $N_\mathrm{max}=N_\mathrm{tot}\cdot m$ where $N_\mathrm{tot}$ is the total number of fluid elements employed in the simulation,
and $m\ge 1$ (we set $m=20$ and $N_\mathrm{tot}=100$ in our tests).
When the buffer is full ($N=N_\mathrm{max}$) we simply move the last $N_\mathrm{max}/2$ buffered cells at the beginning
of the buffer and we set $N=N_\mathrm{max}/2$, in order to re-use the last $N_\mathrm{max}/2$ cells stored.

We provide a pseudocode (Algorithm \ref{pseudo:buffer}) in order to clarify the above description. Note that the pseudocode refers to a given fluid element
(called test element) with initial conditions $\mathbf{x}$ (line 1) that is compared to all the particles in the buffer via the loop $j=N\to1$.
The aim of this algorithm is to find the final conditions $\mathbf{\hat x}$ for the test particle by comparison, (line 15) or, if comparison fails, via 
a call to the solver (line 20). In the algorithm we first check the condition for the time-step $\dd t$ (line 4) that
is written in order to be false when the time-steps of the test particle and the \jth buffer particle are different (line 6) 
to avoid useless calculations (i.e. skip to the next buffer particle). Then it loops over the set of particles 
to compare the initial conditions of the test particle and of the \jth buffer particle (line 8).
In this case we also write the conditions in order to save CPU time by exiting from the initial conditions loop (line 11). 
If both of these tests (time-step and initial conditions)
are false, the variable \emph{found} remains set to true (line 14), and in this case the \jth buffer particle is similar to the test particle: 
we load then the final conditions from the buffer particle (line 15) and we break the loop over the buffer (line 16). 
If, after the loop over the buffer, the variable \emph{found} (line 19) is still false
we need to calculate the final conditions in the standard way (via solver, line 20), we store the new particle in the buffer (lines 21 and 22),
and we increase the size of the buffer (line 24). 

In the tests proposed in Sect.\ref{sect:test}, $70\%$ of the calls to the solver are saved
with a speed-up of $\times5$ and a very small error on the chemical evolution. We also found, as expected, that the number of skipped
calls to the solver decreases during the simulation, because when the shock is fully developed there are less similar fluid elements than at the beginning,
but, more in general, the total number of evolved fluid elements loaded from the buffer (and hence not calculated with the solver) 
depends on the features of the simulated environment. Note that a large buffer ($N_\mathrm{max}\gg N_\mathrm{cell}$) reduces the rate
of efficiency decreasing, although a large buffer needs a non-negligible computational time to be scanned.
{The latter considerations suggest to use this technique when the probability of having similar fluid elements is not very low. However, it depends on the chemical network employed: if the final abundances are weakly dependent on the initial conditions one can choose a less thight similarity threshold, and in this case - during the simulation - the number of similar fluid elements found would be higher, thus obtaining a larger speed-up. Conversely, if the final abundances are affected by small perturbations of the initial conditions this method will probably give a small speed-up when the whole simulations will be fully developed.}

\subsection{Other strategies}

Evaluating the rate coefficients is another source of CPU-consumption. 
Often the ODE for the temperature is coupled with the others ODEs through cooling and/or heating and trough the adiabatic
index ($\gamma$) that depends on the mass fractions of the species.
For this reason the rate coefficients must be evaluated at each solver's step, and usually these coefficients contain intrinsic 
functions as $\exp()$ and $\log()$ that are expensive in terms of computer time. 
One way to solve this problem (while keeping the same accuracy) is to tabulate the rates and to interpolate them during the simulation
when required, obtaining a speed-up up to a $\times5$ depending on the reaction rates used.

The last solver-related issue concerns the function called by the solver that contains the differential equations 
surmised by Eqn.(\ref{eqn:ODE}). These equations can be either explicitly written equation-by-equation in the code, or 
called by using a loop. The latter approach produces a more compact code which is easy to handle and modify, 
and it is faster during compilation (especially for very large networks), 
but unfortunately it is less efficient at runtime. In the tests proposed here we have preferred for the sake of simplicity 
the loop approach, even if explicitly writing the equations would give an additional speed-up.

\section{The reduction methods}\label{sect:reduction_methods}
In this Section we review those reduction strategies that are more directly related to the properties and the features of the network itself. 
We briefly discuss the most used methods for
reducing the workload associated with the chemistry, and we also introduce a new approach based on the topology  of the network.
The need to reduce the astrochemical networks arises from the huge computational time required for astrophysical simulations. Common practices, based on arbitrary cutting, are often used to deal with this problem, e.g. those based on chemical or physical considerations \citep{Ruffle02,Wiebe2003}. As already stated in the Introduction, this approach is reasonable for very small networks where the complexity of the connections among the species is clear enough to decide what to remove, {or when one can recognize some necessary species for the given chemical network}.

There are several methods aiming at reducing the computational cost of the chemistry. A common practice is to pre-calculate the models 
with some one-zone chemical code in order to create a database of model evolutions, and then, when needed, to interpolate the final conditions during 
a large simulation (e.g. \citealt{Bruderer2009}). This method works fine for a small number of parameters, e.g. the temperature and the 
initial number densities of the species involved. When the number of species grows, the dimensionality
of the database increases thus making the database huge and difficult to interpolate.

The latter problem can be solved with an artificial neural network (ANN) approach which works
similarly to an interpolator, but can handle a larger number of free parameters \citep{Grassi2011b}. 
Unfortunately an ANN needs a lot of fine-tuning to reproduce the results
of the database, and for this reason most of the times it cannot be included in large simulations as an interpolator.

Another class of reduction methods is based on linear algebra. (i) The lumping method \citep{Okino1998} allows one to reduce
the dimensionality of the problem by making some linear combination of the species involved. This method has been developed 
for small systems and for this reason cannot be applied to complex networks.
(ii) Also the singular value decomposition (SVD) and the principal component analysis (PCA) are aimed at reducing 
the dimensionality of the ODE system using a transformation matrix that allows to redefine the coordinates of the space of the parameters
(in our case temperature and all the numerical abundances) in order to define a new space where
the dimensionality of the problem is smaller (see e.g. \citealt{Vajda1985}, \citealt{Liu2010}, \citealt{SchultzPhdThesis}). The main drawback is that computing this transformation matrix 
has a non-negligible computational cost and it could easily result in a computational overhead.

The overview of these methods is presented in Tab.\ref{tab:overview}, which includes their \emph{pros} and the \emph{cons} features in terms of usage.

\begin{table}
	\caption{Overview of the methods discussed in this Paper, where \emph{d.o.f.} means degrees of freedom, \emph{transf.} transformation, and
		\emph{infos} is for informations on the structure of the chemical network. See text for further details. \label{tab:overview}}
	\centering
	\begin{tabular}{l|l|l}
		\hline
		Method & pros & cons\\	
		\hline
		Direct & exact & very slow \\
		Interp. & fast & few d.o.f. \\
		ANN & fast, many d.o.f. & hard fine-tuning \\
		User-based & fast & arbitrary \\
		Lumping & fast & small-systems \\
		SVD, PCA & fast & transf. overhead \\
		Buffer & fast & efficiency decreases \\
		Flux-based & fast, infos & DLSODES overhead\\
		Topology & fast, infos & misses initial cond.\\
		\hline
	\end{tabular}
\end{table}

We tackle this computational problem by employing instead a more robust technique known as the flux-method 
presented in \citet{Tupper2002} and \citet{Grassi2012}.
It is an \emph{on the fly} procedure aimed at determining the less active reactions and then reducing the number of 
terms in the RHS of Eqn.(\ref{eqn:ODE}). The tests proposed in \citet{Grassi2012} showed large speed-ups ranging
from $\times2$ to $\times10$ depending on the network employed. It is worth noting that this method has been developed for the 
solver DVODE, while when applied to DLSODES solver it generates an overhead since it interferes with the solver's subroutines 
aimed at handling the sparsity of the Jacobian (Tupper, private communication).

A new \emph{a priori} method based on the connectivity features of a network (i.e. on topology) is therefore 
presented in the next Section,
and some of its key aspects and results will be discussed there.

\subsection{The topology of astrochemical networks}
An astrochemical network can be viewed as a \emph{directed} graph where the vertexes (or nodes) are the species of the chemical system, while
the edges (or links) are determined by reactions between species. A reaction as $\mA+\mB\to\mC$ is represented with three nodes (namely A, B, C)
and two edges ($\mA\to\mC$ and $\mB\to\mC$). When we consider astrophysical networks the number of species/vertexes spans from a few tens to
more than several hundreds, while reactions/edges can be more than five thousands. This large number of items becomes complicated
to be displayed as a graph and, as a consequence, the topological features are usually not evident at first sight.
Different methods have been proposed through the years to explore the properties of the networks as summarized by \citet{Jolley2010,Jolley2012}.
A widely used method consists in measuring the degree of the nodes that comprise the network, where degree is defined as the sum of the
connection of a node to its neighbourhoods. This quantity shows some interesting properties when applied to real examples. 
One of the most intriguing is the power-law distribution of the probability of finding a node with a given number of connections.
More explicitly the probability of finding a node with a degree $d$ decreases as $P(d)\propto d^{-\gamma}$ (where $\gamma>0$ is a parameter), 
it means that
finding a node with many connections is less probable that finding a less connected one.

This property is true for \emph{scale-free} networks, that are considerably robust and stable to random node removing (e.g. \citealt{Barabasi1999}).
In particular when the network is robust it allows to remove some nodes without damaging
the global stability of the network because the largest part of the information flows through the so called \emph{hubs}
that are the most connected elements of the system. When we deal with astrophysical networks we are interested in finding 
stable structures from a ``noisy background'' represented by the less active nodes or, as in the scheme just depicted, we want to know
what node can be removed. 

Even if astrochemical networks are not \emph{scale-free}, since they have an exponential degree distribution 
instead of a power law \citep{Jolley2010}, we can determine which are the sub-structures
that can be deleted by using some ranking techniques.
The degree of a node can be a good way to determine whether a node is important or not,
and then if it could be removed or not; when we decide to use the degree as a criterion
we assume that it is very probable that a large part of the 
information will flow through very connected nodes instead of flowing through the less connected. It is evident that
removing vertexes with many connections (\emph{hubs}) could seriously damage the whole network because of their crucial role in the global
network activity. {However, a method based on the degree could lead to wrong results, for example when a node with few connections acts as a bridge between two largely connected sub-networks: in this case its removal will drastically reduce the information exchange between the two sub-networks, and it could determine improper final abundances.}

{To avoid the latter issue we introduce} another approach to rank the most important species, that is to calculate how much each node is linked to nodes that are highly connected to the rest of the network. 
This method results from the assumption that the most connected nodes carry the largest
part of the global network information and in particular we choose that being the neighbour of a very connected node increases the probability
of becoming active during the network evolution.
This fact can become more clear when we take a social network as an example: in this case a person (a node) can easily reach critical information
if it is connected to a few individuals that massively participate to the information exchange (\emph{hubs}), rather than being in contact with a large number of poorly connected persons.
 
We call this ``second degree'' and it can be represented as the sum of the degrees of the nodes close
to a given node (i.e. its neighbours). For the \ith node is
\begin{equation}\label{deg2_def}
	^{(2)}d_i=\sum_{j\in N_i}d_j\,,
\end{equation}
where $^{(2)}d_i$ is the second degree of the \ith node, $N_i$ is the set of the neighbours of the \ith node, and
$d_j$ is the degree of the node $j$.

In this way the second degree becomes a parameter that is more powerful than the degree itself, because it takes into account
not only the activity of a given node, but also the activity of its close neighbours. In principle we can also calculate
the \nth degree by iteratively applying the idea above for $n\to\infty$
\begin{equation}
	^{(n)}d_i=\sum_{j\in N_i}{}^{(n-1)}d_j\,.
\end{equation}
This is similar to the Google PageRank \citep{Page1999} that assigns a numerical weighting 
to each web page determining its relative importance. Both algorithms have an iterative approach aimed at 
finding the probability distribution that represents the likelihood that a given species participates to the 
global network activity, or a person randomly clicking on various links will arrive at a particular page.

In this first analysis, to compute the second degree we consider the graph as \emph{undirected}, since to determine
the activity of a given node we have to include the edges that leave that node, which represent the destruction reactions,
and the edges pointing at that node, being reactions which form the species represented by the given node.
Under this assumption the activity of the node is determined by both the reactions that form and destroy the corresponding species,
hence the need to consider the graph as undirected.

There are three approaches to calculate the second degree and the difference between them is determined by the weight
assumed during the calculation of the degree $d_i$ in the equation 
\begin{equation}\label{eqn:degree_w}
	d_i=\sum_{j\in N_i}w_jd_j,\
\end{equation}
where $w_j$ is the weight. The first method to obtain $d_i$ is simply (i) to count the number of connections
that reach or leave a node assuming $w_j=1$. This method allows to compute the second degree without considering 
the time evolution, but it neglects whether the reactions carrying information to the node are important or not. 
(ii) To improve on that choice, one can change the weight to $w_j = k_{ij}$ where $k_{ij}$ is the rate coefficient
of the reaction that involves the \ith node as product and the \jth node as a reactant. 
This is more accurate than the first approach, but it 
cannot be calculated once and for all since $k_{ij}$ is temperature-dependent. This method also 
ignores the fact that at a given time a reaction could not be active because the abundance of the reactants is zero.
(iii) The latter issue can be avoided by considering $w_j=F_{ir}$ where $F_{ij}$ is the flux of a reaction
calculated as $F_{ij}=k_{ij}\,n_j\,\prod_r n_r$ where $n_j$ is the abundance of the neighbour reacting species,
and the product runs over the other reactants excluding $n_j$. This calculation can be only performed at runtime, because
one needs to know the abundances of the species during the simulation, and moreover, the rate coefficient $k_{ij}$
depends on the temperature, which changes during the evolution of the model.

In this paper we shall show the results only for the first method that can be used as a good estimator to determine the most important species,
thereby allowing to eliminate from the network the less active ones before calculation. The other two methods have more general validity and accuracy 
but require evaluation during runtime, which is a significant drawback from the computational standpoint.

Once we have ranked the species in the network we can choose a criterion to remove the less important ones,
and to choose this we want to make sure that we do not remove species with high abundances.
To cope with this risk we introduce a second degree threshold $d_n$ based on species densities
\begin{equation}
	^{(2)}d_n = \mathrm{min}\left[{}^{(2)}d_i\right] \forall\,i\,|\,n_i>n_t\,.
\end{equation}
that depends on the lowest of the second degrees ${}^{(2)}d_i$ of the species with a number density $n_i$ 
greater than a user-defined density threshold $n_t$.
We also define a second threshold $d_f$ to keep important species when the first threshold can be excessively constraining 
(e.g. initial monoatomic gas), namely
\begin{equation}
	^{(2)}d_f = f\cdot\max\left[{}^{(2)}d_i\right]\,.
\end{equation}
which is determined by a certain fraction $f\le1$ of the maximum second degree.

We decide to cut the species that do not satisfy both criteria at the same time, hence the final threshold becomes
\begin{equation}
	^{(2)}d_t = \min\left[^{(2)}d_n,{}^{(2)}d_f\right]\,,
\end{equation}
which gives the final criteria ${}^{(2)}d_i>{}^{(2)}d_t$ employed to decide whether or not the \ith species is kept in our calculation.
In the tests we present in Sect.\ref{subsect:test_topology} we use $f=0.6$ and $n_t=10^{-5}$.

{This procedure for determining the threshold is not completely ``automatic'', and a certain knowledge of the characteristics of the chemical network employed is needed (e.g. when the network is employed to study the sulphur chemistry the threshold will be chosen such to avoid cutting many of the sulphur-bearing species). However, we have proposed here a threshold that is a good compromise between an ``automatic'' and a completely user-based one, keeping in mind that the key-feature of the topological method resides on the possibility of ranking the species by their importance.}

\section{Test cases}\label{sect:test}
We propose here tests involving a selection of the aforementioned methods, employing the OSU reactions database 
as in \citet{Wakelam2008} - without PAHs - coupled to a simple 1D Lagrangian code aimed at simulating a Sedov-like gas shock \citep{Bodenheimer2006}, {see their Sect. 6.2 and 10.3 for equations and details}.

{The aim of this code is to provide a numerical benchmark to test the efficiency of our reduction methods, since we have introduced a chemical network that is developed mainly for dark cloud-like environments. Moreover, in our case we note that we had to extrapolate some of the reactions to the higher temperatures that were reached during the shock. While the accuracy of this procedure is somewhat uncertain, we note that the case considered here merely represents a stress-test for the reduction schemes we propose. We use this set of chemical reactions because it represents one of the largest astrochemical network available and it is computationally challenging when coupled to a shock code as the one proposed by \citet{Bodenheimer2006}.}

The model is a gas composed of a set of spherical shells representing an inner region with initial conditions $T_\mathrm{in}=10^4$ K 
and $\rho_\mathrm{in}=10^{-20}$ g/cm$^3$, and another set of shells for the outer region with 
$T_\mathrm{out}=10$ K and $\rho_\mathrm{out}=10^{-21}$ g/cm$^3$ and radius $R = 1 pc$. 
The number of the inner shells is $N_\mathrm{in}=40$, while the number of the outer shells is $N_\mathrm{in}=60$, giving
$N_\mathrm{tot}=100$. We should note here that the results found for the tests presented in this Section are still valid for larger systems, for systems based on an Eulerian approach, and for 3D hydrodynamics.

The initial abundances are the same as the EA2 model of \citet{Wakelam2008} (i.e. solar metallicity initial conditions) and 
we let the system evolve for $10^7$ yr. All the tests presented in this paper employ the DLSODES solver 
with an absolute and relative tolerance of $10^{-40}$ and $10^{-12}$ respectively 
(which are accurate enough compared to the values of the number densities usually employed in ISM simulation), 
and an internally-generated Jacobian (option MF=222, more
details in \citealt{Hindmarsh83}). Note also that the code has largely been optimized and compiled with 
Intel$^\circledR$ Fortran Composer XE 2011 Update 6 using a standard set of optimization flags.

It is important to remark that since the scope of these tests is to determine the computational efficiency of the methods proposed, 
we use the OSU database unchanged (\osu, 4431 reactions and 452 species) and,
moreover, we do not include any cooling or heating. Under the latter assumption the hydrodynamics of the shock is not affected by the chemical evolution.
However, the tests discussed here represent - computationally speaking - a typical ISM scenario with a large chemical network. We plan to describe
the coupling between hydrodynamics, chemistry and cooling together with reduction methods in a forthcoming paper.

{To ensure the validity of the results obtained we have independently evolved the chemistry from the initial conditions both for an inner and an outer fluid element. These tests are performed using a one-zone pseudo time-dependent code \citep{Grassi2011,Carelli2013} employing the same chemical network of the 1D tests reported here. As expected the results found for both codes are in complete agreement}.

\begin{figure}
\begin{center}
	\includegraphics[width=.45\textwidth, angle=0]{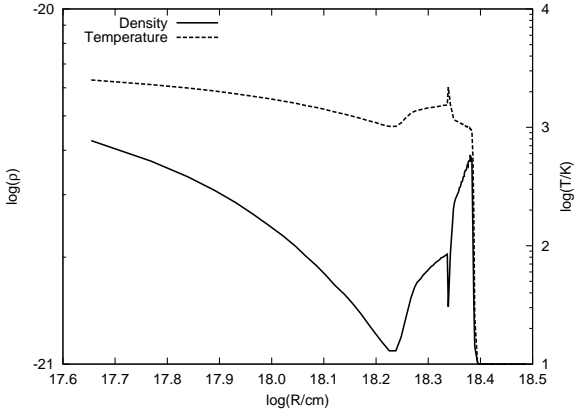}
	\caption{Radial profiles of the density (solid) and temperature (dashed) when the shock is fully developed, with $\rho$ in g cm$^{-3}$.}
	\label{fig:density}
\end{center}
\end{figure}

{The radial profiles of the density and temperature of the developed shock are shown in Fig.\ref{fig:density}. The shock is propagating left to right, and we distinguish the freely-expanding ejecta extend out to $\log[R/$cm$]\simeq18.2$. From 18.2-18.3 we have the reverse shock and the shocked ejecta, while the density jump represents the contact discontinuity. Finally, the shocked ambient medium is 18.35-18.4.}

\begin{figure}
\begin{center}
	\includegraphics[width=.45\textwidth, angle=0]{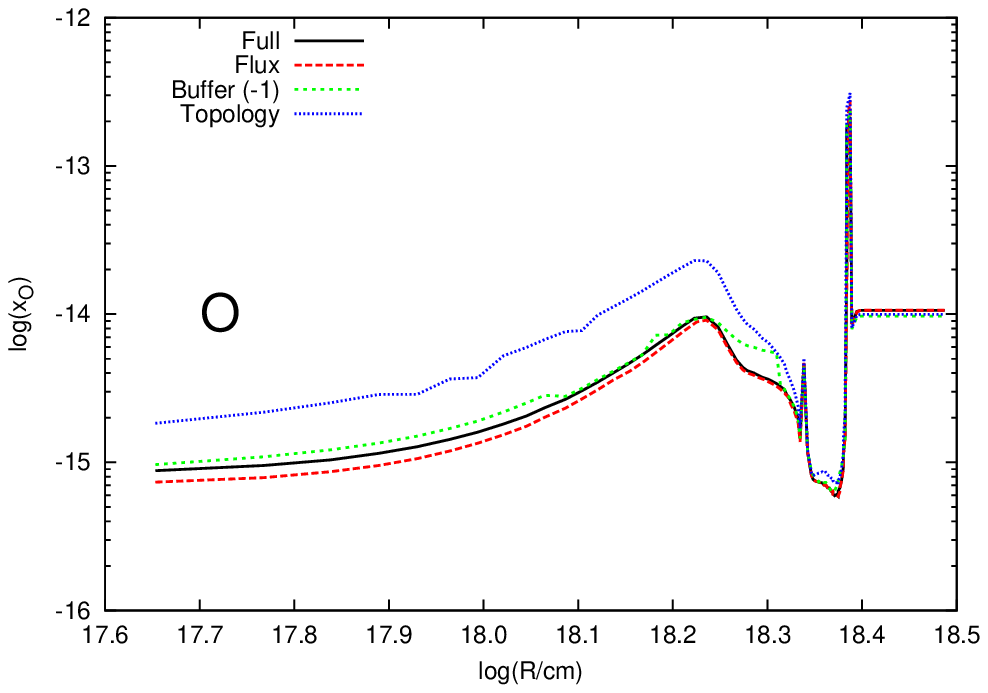}
	\includegraphics[width=.45\textwidth, angle=0]{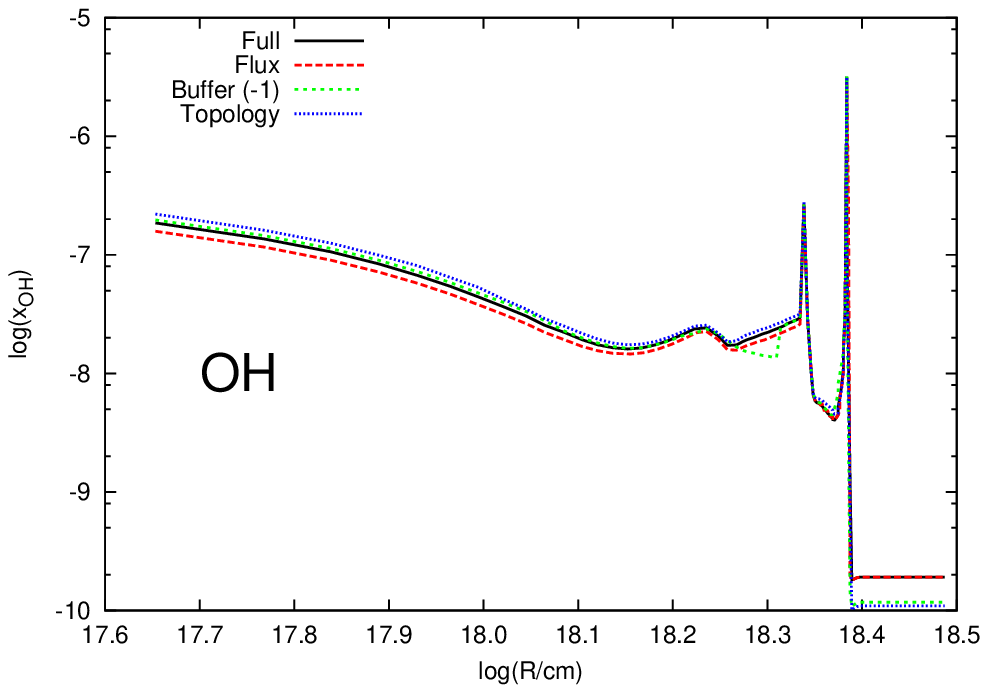}
	\caption{Radial profiles of the fractional abundances of oxygen (top) and OH molecule (bottom) when the shock is fully developed. The black solid line represents the profile of the model evolved without any reduction methods, that is compared with the profile of the \emph{flux} (long-dashed, red), \emph{buffer} (dashed, green), and \emph{topology} (dotted, blue) methods. The profile of the \emph{Buffer} method with $\epsilon=10^{-5}$ is not displayed since it overlaps the \emph{full} model profile. For colour figures see the online version.}
	\label{fig:pro_O}\label{fig:pro_OH}
\end{center}
\end{figure}

\begin{figure}
\begin{center}
	\includegraphics[width=.45\textwidth, angle=0]{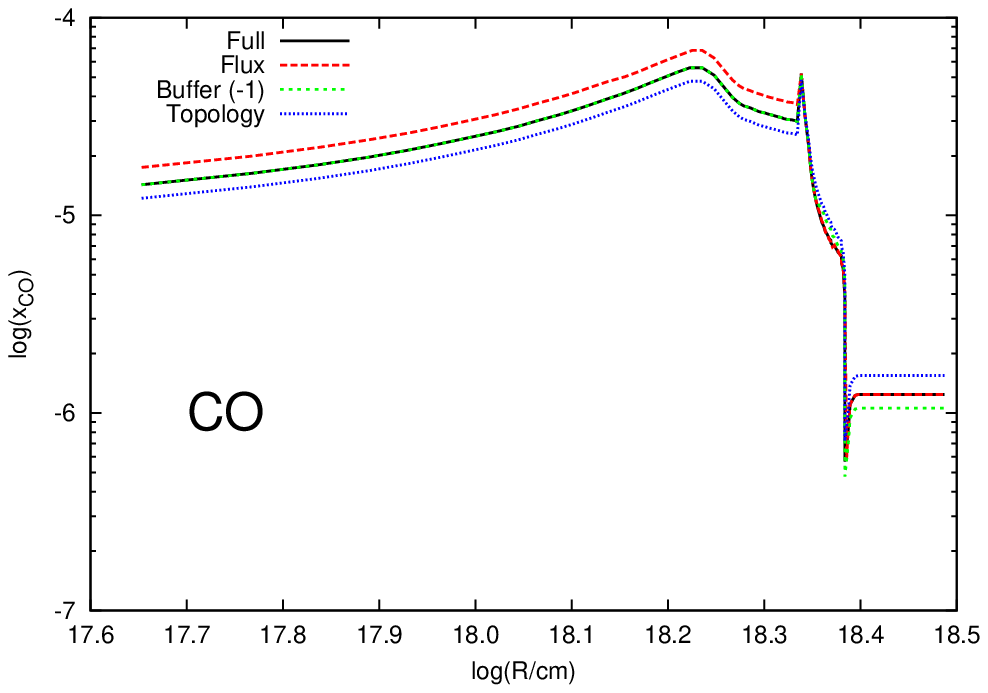}
	\includegraphics[width=.45\textwidth, angle=0]{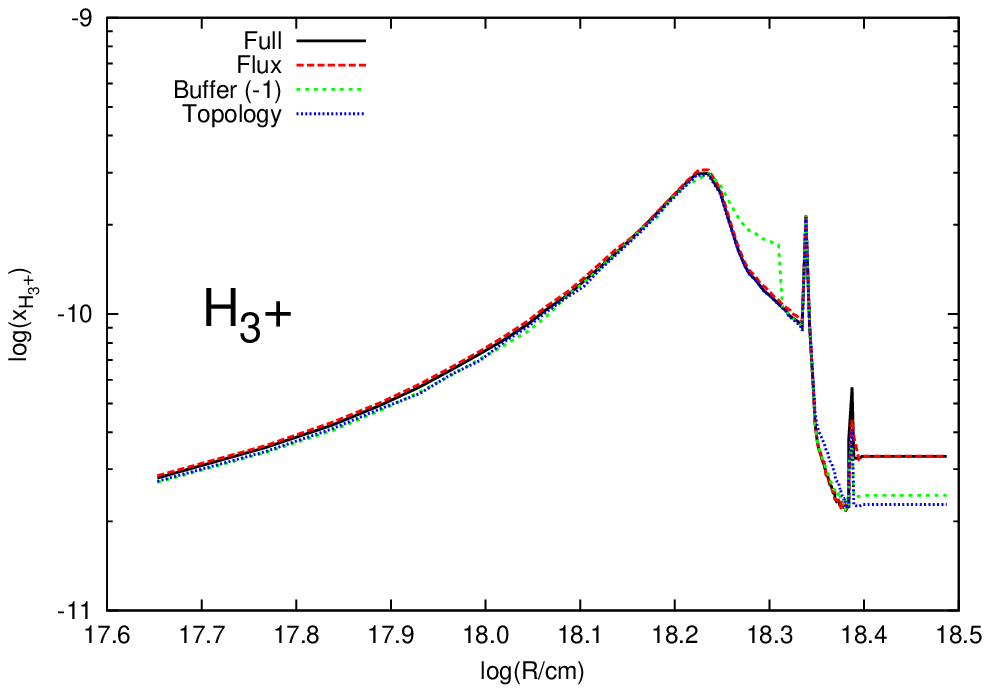}
	\caption{Radial profiles of the fractional abundances of CO (top) and H$_3^+$ molecule (bottom) when the shock is fully developed. The black solid line represents the profile of the model evolved without any reduction methods, that is compared with the profile of the \emph{flux} (long-dashed, red), \emph{buffer} (dashed, green), and \emph{topology} (dotted, blue) methods. The profile of the \emph{Buffer} method with $\epsilon=10^{-5}$ is not displayed since it overlaps the \emph{full} model profile. For colour figures see the online version.}
	\label{fig:pro_CO}\label{fig:pro_H3p}
\end{center}
\end{figure}

The methods are all compared with the \emph{full} evolution, i.e. without any reduction,
{and in particular we show the radial profile of the fractional abundances of oxygen and of some sample molecules as CO, CH, and H$_3^+$ (see Figs.\ref{fig:pro_O} and \ref{fig:pro_CO}) when the shock is fully developed. We show the profiles of the model without reduction (solid line labelled \emph{full}) compared with \emph{flux}, \emph{buffer} with $\epsilon=0.1$, and with \emph{topology} radial profiles. We have omitted the radial profile of \emph{buffer} method with the smaller threshold since the differences with the \emph{full} model profile are not visible. The worst case is represented by the \emph{topology} method which achieves the worst results as indicated by Fig.\ref{fig:pro_O} (top), while the \emph{buffer} method shows good results except for the region close to $log(R/cm)=18.3$ which clearly differs from the \emph{full} model profile. This separation is not present when the threshold is $\epsilon=10^{-5}$. Conversely, the \emph{flux} method yields the best results.}

{Comparing the density and temperature profiles (Fig.\ref{fig:density}) with the abundances of the species (Fig.\ref{fig:pro_O} and \ref{fig:pro_CO}) we recognize that the formation of the selected species is influenced by the temperature discontinuities corresponding to the spikes in the abundances profiles. We roughly identify three regions in the abundances profiles with different chemical behaviour: namely (i) the undisturbed medium ($\log[R/$cm$]\lesssim18.2$) where the chemical abundances have reached the steady state for an environment with $T\simeq10$ K and density \mbox{$\rho\simeq10^{-21}$ g cm$^{-3}$} (both almost constant during the shock evolution),  then (ii) the region crossed by the shock ($18\lesssim\log[R/$cm$]\lesssim18.2$) that shows complex features including the spikes previously described, and finally (iii) the inner region that has evolved to the steady state analogously to the outer region, but with an higher temperature $T\gtrsim10^3$ K that decreases during the evolution from the initial temperature of $T=10^4$ K. }

\begin{figure}
\begin{center}
	\includegraphics[width=.45\textwidth, angle=0]{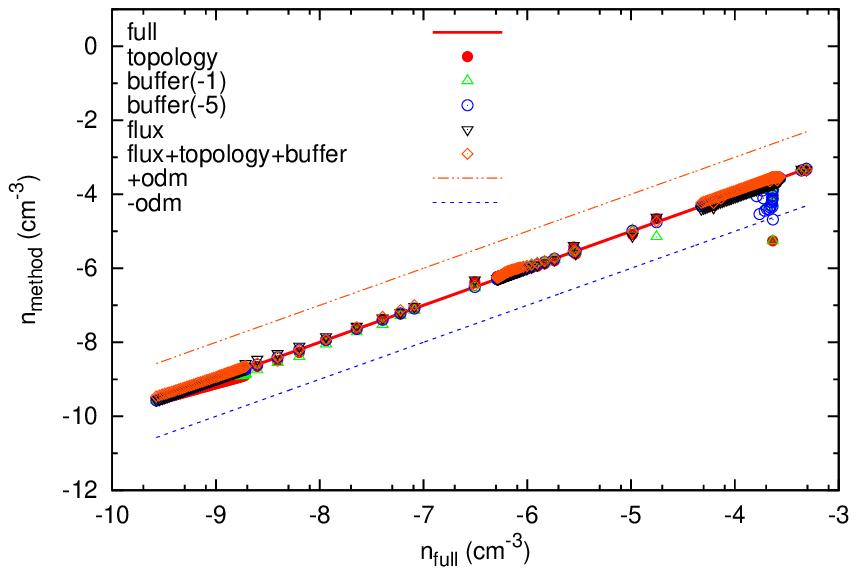}
	\includegraphics[width=.45\textwidth, angle=0]{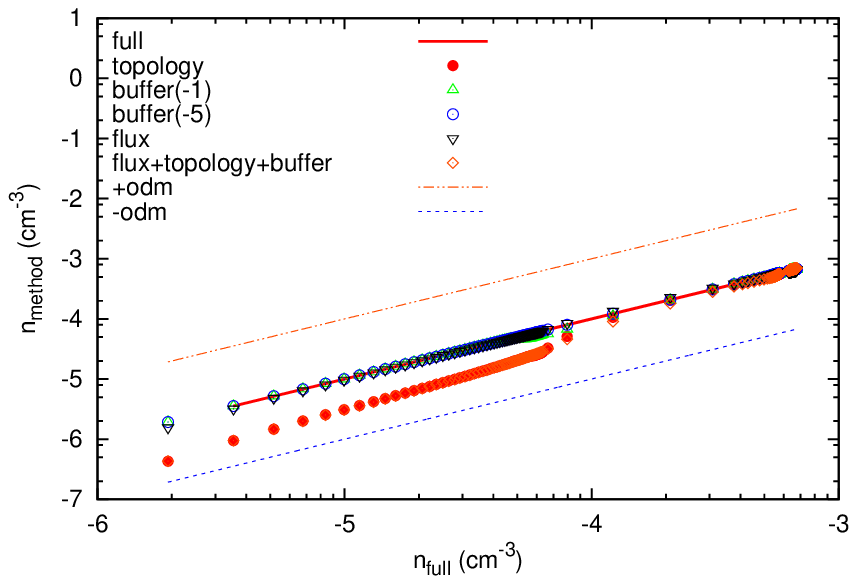}
	\caption{Comparison of the \emph{full} model with the different reduction methods for OH (top) and H (bottom). As indicated in the key
		the methods are \emph{topology} (dots), \emph{buffer} with $\epsilon=10^{-1}$ (triangles up) and $\epsilon=10^{-5}$ (open circles),
		\emph{flux} (triangles down), and hybrid (diamonds). For colour figures see the online version.
	}
	\label{fig:xyOH}\label{fig:xyH}
\end{center}
\end{figure}

Moreover, to show the accuracy of the different methods we have included the plot of the abundances of a given species in all the gas shells for each reduction method compared to the same shell in the \emph{full} model (see Figs.\ref{fig:xyCp} and \ref{fig:xyOH}). More specifically these plots represent the accuracy in reproducing the chemical behaviour of 
the gas during the evolution of the shock in all the shells for a given species. If a method reproduces exactly the chemical evolution
of the \emph{full} model all the points will lie along a straight line (i.e. $y=x$), while the distance from the line increases if the method fails to reproduce
the original model. These plots represents the dispersion of the values obtained for a given method ($n_\mathrm{method}$) compared to the values of the full model $n_\mathrm{full}$. We have also included two lines representing the range of fluctuations by one order of magnitude in each direction 
(i.e. $y=10\,x$ and $y=0.1\,x$ labelled +odm and -odm respectively).

We found that by coupling together \emph{buffer} and \emph{flux} methods gives the best results both in terms of robustness and of speed-up of the calculations.

\subsection{Comparison of solvers}\label{sect:test_solver}
The characteristics of the chemical network play a key role to determine what is the best numerical solver. 
In this framework we propose a 1D shock model test mainly focused on Jacobian sparsity, 
 using DVODE and DLSODES solvers \citep{Hindmarsh83, Hindmarsh2005}.
It is important to remark that employing an internally generated or a user-provided Jacobian will affect the performance 
of the solver, since building the Jacobian has a non-negligible cost for the solver; 
however for the sake of simplicity we use this latter method in all our tests.

The Jacobian associated with the network of the OSU database is extremely sparse ($\sim94\%$), 
which is very common for astrochemical networks. For this reason in the shock test framework 
the numerical solver DLSODES outperforms the more general DVODE solver achieving approximately a $\times100$ speed-up.

\subsection{Buffer method}\label{subsect:test_buffer}
We test the \emph{buffer} method with two thresholds, namely $\epsilon=10^{-1}$ and $\epsilon=10^{-5}$, the latter being a finer 
approximation. We found good results especially for the $10^{-5}$ threshold which reproduces the full model almost exactly,
but, as expected, when increasing the threshold value ($\epsilon=10^{-1}$) the method starts to be less accurate in reproducing 
the original data, but the error always remains below one order of magnitude (see Fig.\ref{fig:xyCp} and Fig.\ref{fig:xyH}).

\subsection{Topological method}\label{subsect:test_topology}
\begin{figure}
\begin{center}
	\includegraphics[width=.45\textwidth, angle=0]{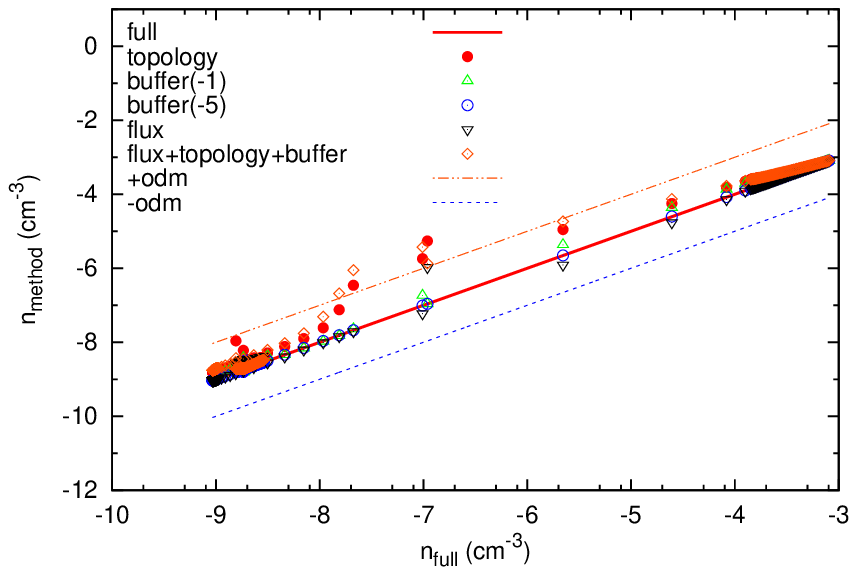}
	\includegraphics[width=.45\textwidth, angle=0]{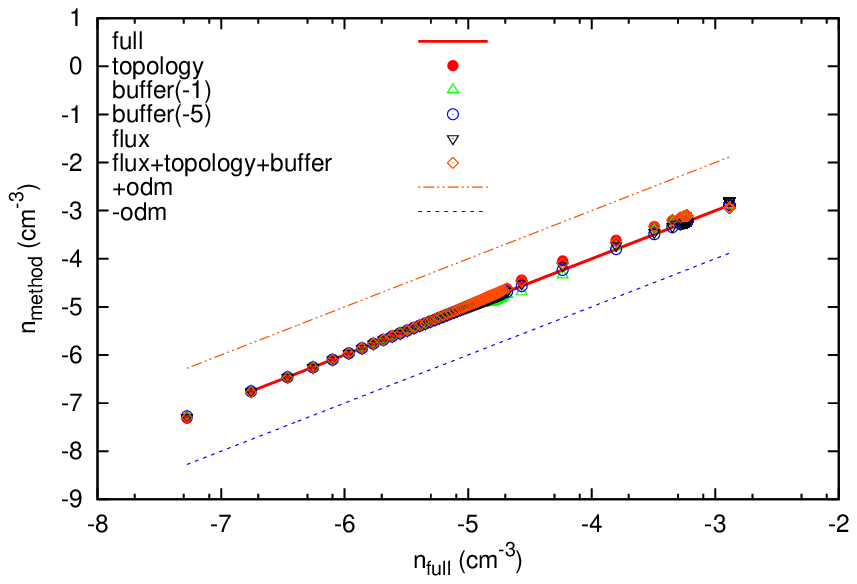}
	\caption{Comparison of the \emph{full} model with the different reduction methods for C$^+$ (top) and CO (bottom). As indicated in the key
		the methods are \emph{topology} (dots), \emph{buffer} with $\epsilon=10^{-1}$ (triangles up) and $\epsilon=10^{-5}$ (open circles),
		\emph{flux} (triangles down), and hybrid (diamonds). For colour figures see the online version.
	}
	\label{fig:xyCp}\label{fig:xyCO}
\end{center}
\end{figure}

The \emph{topological} method reproduces the \emph{full} model with a good accuracy even if some dispersion is observed for the C$^+$ and H species 
(Fig.\ref{fig:xyCp}, top and Fig.\ref{fig:xyH}, bottom).  The threshold we chose selects 2369 reactions over 4431, hence it removes 
almost 50\% of the network connections. As stated in the previous Section, the reduction based on the topological method has a 
probabilistic meaning and its best property lies in the capability of giving information on which species/reactions are important or not. 
For instance, our tests show that the most important \emph{hubs} (the most connected species) are, in that order, free electrons, H, H$_2$, 
He$^+$, C, H$^+$, O and C$^+$, as expected. This topological approach finds that species like Cl, Mg, Fe, and their ions, but also MgH, 
and HF and many C-chains species, are not so important within this network, thus many reactions involving He$^+$ and C$^+$ are neglected. 
The latter can explain the fluctuation of C$^+$ species as shown in Fig.\ref{fig:xyCp}. We want to underline 
once again that, unlike \emph{flux} or \emph{buffer}, the 
topological method is mainly aimed at giving \emph{a priori} chemical information on the network rather than strongly reducing the
number of the reactions, or the species.
A more robust reduction technique based on topology has already been discussed: using the flux as the weight in Eqn.(\ref{eqn:degree_w}) will provide a more accurate analysis. However, this method is not \emph{a priori}, and for this reason it could generate a 
large overhead when the reduction is less effective. Moreover, from a computational point of view, this technique is closer to the \emph{flux} method, 
since the weights are exactly the fluxes and the two methods are largely comparable. 
The only advantage is the amount of topological information provided during the system evolution.
Due to these considerations, it seems likely that flux-based reduction techniques are preferable when a robust reduction
method is required, even if the error associated with the \emph{topology} method is generally less than one order of magnitude.
Depending on the desired accuracy, the method can still be useful in some specific applications.

\subsection{The flux-based method}
For the \emph{flux} method presented here we assume that the length of the
hydrodynamical time-step is equal to the length of the macro-step, 
which is a realistic approximation since temperature and total density will remain constant during a hydrodynamical time-step. 
Under this assumption the evaluation of the fluxes is made at the beginning of each hydrodynamical time-step for each gas shell. 
This allows one to couple the DLSODES solver with the flux method. The results (Fig.\ref{fig:xyCp} and Fig.\ref{fig:xyH}) are in good agreement with the full solution and can be further improved
using a more accurate definition of macro-step (see \citealt{Grassi2012} for further details).  

\subsection{Hybrid methods}
The last method discussed here is a mixing between \emph{buffer} ($\epsilon=10^{-5}$), \emph{flux}, and \emph{topology}. The accuracy of this
hybrid method is determined by the least accurate approach as we can see in Fig.\ref{fig:xyCp} (top), where topology fails, and also in Fig.\ref{fig:xyH} (bottom) where topology is not as accurate as the other methods.

The normalized CPU-times of the different methods are reported in Table \ref{tab:speedup}: as expected hybrid methods are
the fastest, but coupling \emph{flux} and \emph{buffer} shows a better accuracy than the coupling of \emph{flux}, \emph{buffer}, and 
\emph{topology}. We also obtain a large speed-up for the coarsest of the two \emph{buffer} methods, while the one with  $\epsilon=10^{-5}$
has the smallest speed-up, although it has the best accuracy among the methods proposed. Finally, the \emph{topology} shows good speed-up
with less accurate results. It is worth noticing that the latter reduction method gives important information about the chemical network
due to its topological approach, and that to really see the effects of such a cut on the global hydrodynamical evolutions we need a test which couples the chemistry to the dynamics via heating and cooling, and via photon diffusion.

\begin{table}
	\caption{Normalized CPU-time measured for different reduction methods. See text for further details. \label{tab:speedup}}
	\centering
	\begin{tabular}{l|l|l}
		\hline
		Method & $t_\mathrm{CPU}$\\	
		\hline
		Full & 1.00\\
		Topology & 0.37\\
		Buffer (-1) & 0.40\\
		Buffer (-5) & 0.61\\
                Flux & 0.50\\
		Flux+Buffer & 0.24\\
		Flux+Buffer+Topology & 0.17\\
		\hline
	\end{tabular}
\end{table}

\begin{table}
	\caption{Speed-up overview for the methods discussed here. Note that the values listed here refer to 
		the ISM model presented in this paper. One can find different speed-ups depending on the chemical 
		network employed, the numerical framework, and some other parameters. However, these values are
		representative for the methods discussed and can be used as reference. \label{tab:fspeedup}}
	\centering
	\begin{tabular}{l|l|l}
		\hline
		Method & Speed-up\\	
		\hline
		Solvers & $\times$100\\
		Reduction methods & up to $\times$10\\
		Buffering & up to $\times$5\\
		Rates tabulation & $\times$5\\
		\hline
	\end{tabular}
\end{table}

\section{Conclusions}\label{sect:conclusions}
In this Paper we have discussed some techniques aimed at reducing the computational cost
of including a chemical network into astrophysical codes that simulate the evolution of the ISM.
We have divided the methods into two classes, namely the \emph{``pure'' computational strategies} that are more
related to computer science, and the \emph{reduction methods} that  are based on the properties and on the structure
of the chemical network. In the first class we discussed the solvers employed, the buffering method, and a brief review on
other strategies, while in the second class we described the flux method and introduced a topology-based method.

To test the different methods we employed a 1D hydrodynamical Sedov-like shock test with a standard OSU (\osu) chemical network,
which is a large chemical network ($>4000$ reactions). 
The methods tested are \emph{buffer}, \emph{topology}, and \emph{flux}, including some hybrid methods that couple the previous ones.
These methods reproduce the results of the \emph{full} model with good speed-ups, except for the topology-based approach which
has a larger error compared to the other methods. The reason is that it is mainly devoted to rank by importance the different species
rather than providing a robust reduction technique as the \emph{flux} method. Nevertheless, the results provided by this method
suggest that the topological approach can lead to an efficient reduction based on the features of the chemical network,
which gives at the same time some critical information about the global properties of the interconnections between the various species
included in the ISM model.

Finally, we found that the most robust method among those examined turns out to be 
the coupling between the \emph{flux} approach and the \emph{buffer} technique,
and this hybrid method also achieves one of the best speed-ups (almost $\times5$) suggested by our tests. The fastest
shock simulation is obtained, as expected, by using \emph{flux}, \emph{buffer}, and \emph{topology} all together, but 
the \emph{topology} affects the global behaviour of this hybrid method producing results that are less accurate than
\emph{flux} and \emph{buffer} methods and their hybrid usage. 

A final overview of the speed-ups obtained from the various methods presented in this paper is given in Table \ref{tab:fspeedup}.

\section*{Acknowledgements}
T.G. acknowledges the financial support from the CINECA and S.B. and D.R.G.S. thank for funding through the DFG priority program `The Physics of the Interstellar Medium' (projects SCHL 1964/1-1). D.R.G.S. thanks for funding via the SFB 963/1 on ``Astrophysical Flow Instabilities and Turbulence''. 
F.A.G. thanks the PRIN2009 project by the Ministry of Research (MIUR).

\bibliographystyle{mn2e}      
\bibliography{mybib}

\begin{thebibliography}{}

\bibitem[\protect\citeauthoryear{{Barabasi}}{{Barabasi}}{1999}]{Barabasi1999}
{Barabasi} A.~L.,  1999, Physica A Statistical Mechanics and its Applications,
  272, 173

\bibitem[\protect\citeauthoryear{Bodenheimer, Laughlin, Rozyczka \&
  Yorke}{Bodenheimer et~al.}{2006}]{Bodenheimer2006}
Bodenheimer P.,  Laughlin G.,  Rozyczka M.,    Yorke H.,  2006, Numerical
  Methods in Astrophysics: An Introduction.
Series in Astronomy and Astrophysics, Taylor \& Francis

\bibitem[\protect\citeauthoryear{{Bruderer}, {Doty} \& {Benz}}{{Bruderer}
  et~al.}{2009}]{Bruderer2009}
{Bruderer} S.,  {Doty} S.~D.,    {Benz} A.~O.,  2009, \apjs, 183, 179

\bibitem[\protect\citeauthoryear{{Carelli}, {Grassi} \& {Gianturco}}{{Carelli}
  et~al.}{2013}]{Carelli2013}
{Carelli} F.,  {Grassi} T.,    {Gianturco} F.~A.,  2013, \aap, 549, A103

\bibitem[\protect\citeauthoryear{{Galli} \& {Palla}}{{Galli} \&
  {Palla}}{1998}]{GalliPalla98}
{Galli} D.,  {Palla} F.,  1998, \aap, 335, 403

\bibitem[\protect\citeauthoryear{{Glover}, {Federrath}, {Mac Low} \&
  {Klessen}}{{Glover} et~al.}{2010}]{Glover2010}
{Glover} S.~C.~O.,  {Federrath} C.,  {Mac Low} M.-M.,    {Klessen} R.~S.,
  2010, \mnras, 404, 2

\bibitem[\protect\citeauthoryear{{Gnedin}, {Tassis} \& {Kravtsov}}{{Gnedin}
  et~al.}{2009}]{Gnedin2009}
{Gnedin} N.~Y.,  {Tassis} K.,    {Kravtsov} A.~V.,  2009, \apj, 697, 55

\bibitem[\protect\citeauthoryear{{Grassi}, {Bovino}, {Gianturco}, {Baiocchi} \&
  {Merlin}}{{Grassi} et~al.}{2012}]{Grassi2012}
{Grassi} T.,  {Bovino} S.,  {Gianturco} F.~A.,  {Baiocchi} P.,    {Merlin} E.,
  2012, \mnras, 425, 1332

\bibitem[\protect\citeauthoryear{{Grassi}, {Krstic}, {Merlin}, {Buonomo},
  {Piovan} \& {Chiosi}}{{Grassi} et~al.}{2011}]{Grassi2011}
{Grassi} T.,  {Krstic} P.,  {Merlin} E.,  {Buonomo} U.,  {Piovan} L.,
  {Chiosi} C.,  2011, \aap, 533, A123

\bibitem[\protect\citeauthoryear{{Grassi}, {Merlin}, {Piovan}, {Buonomo} \&
  {Chiosi}}{{Grassi} et~al.}{2011}]{Grassi2011b}
{Grassi} T.,  {Merlin} E.,  {Piovan} L.,  {Buonomo} U.,    {Chiosi} C.,  2011,
  ArXiv/1103.0509

\bibitem[\protect\citeauthoryear{Hindmarsh}{Hindmarsh}{1983}]{Hindmarsh83}
Hindmarsh A.~C.,  1983, IMACS Transactions on Scientific Computation, 1, 55

\bibitem[\protect\citeauthoryear{Hindmarsh, Brown, Grant, Lee, Serban, Shumaker
  \& Woodward}{Hindmarsh et~al.}{2005}]{Hindmarsh2005}
Hindmarsh A.~C.,  Brown P.~N.,  Grant K.~E.,  Lee S.~L.,  Serban R.,  Shumaker
  D.~E.,    Woodward C.~S.,  2005, ACM Trans. Math. Softw., 31, 363

\bibitem[\protect\citeauthoryear{{Jappsen}, {Klessen}, {Larson}, {Li} \& {Mac
  Low}}{{Jappsen} et~al.}{2005}]{Jappsen2005}
{Jappsen} A.-K.,  {Klessen} R.~S.,  {Larson} R.~B.,  {Li} Y.,    {Mac Low}
  M.-M.,  2005, \aap, 435, 611

\bibitem[\protect\citeauthoryear{{Jolley} \& {Douglas}}{{Jolley} \&
  {Douglas}}{2012}]{Jolley2012}
{Jolley} C.,  {Douglas} T.,  2012, Astrobiology, 12, 29

\bibitem[\protect\citeauthoryear{{Jolley} \& {Douglas}}{{Jolley} \&
  {Douglas}}{2010}]{Jolley2010}
{Jolley} C.~C.,  {Douglas} T.,  2010, \apj, 722, 1921

\bibitem[\protect\citeauthoryear{Liu \& liang Lin}{Liu \& liang
  Lin}{2010}]{Liu2010}
Liu Y.-C.,  liang Lin C.,  2010, in SICE Annual Conference 2010, Proceedings of
  Model reduction of biochemical networks.
pp 3213 --3218

\bibitem[\protect\citeauthoryear{{Meijerink} \& {Spaans}}{{Meijerink} \&
  {Spaans}}{2005}]{Meijerink2005}
{Meijerink} R.,  {Spaans} M.,  2005, \aap, 436, 397

\bibitem[\protect\citeauthoryear{{Merlin} \& {Chiosi}}{{Merlin} \&
  {Chiosi}}{2007}]{MerlinChiosi07}
{Merlin} E.,  {Chiosi} C.,  2007, \aap, 473, 733

\bibitem[\protect\citeauthoryear{{Nejad}}{{Nejad}}{2005}]{Nejad2005}
{Nejad} L.~A.~M.,  2005, \apss, 299, 1

\bibitem[\protect\citeauthoryear{{Nelson} \& {Langer}}{{Nelson} \&
  {Langer}}{1999}]{Nelson1999}
{Nelson} R.~P.,  {Langer} W.~D.,  1999, \apj, 524, 923

\bibitem[\protect\citeauthoryear{Okino \& Mavrovouniotis}{Okino \&
  Mavrovouniotis}{1998}]{Okino1998}
Okino M.~S.,  Mavrovouniotis M.~L.,  1998, Chemical Reviews, 98, 391

\bibitem[\protect\citeauthoryear{{Omukai}, {Tsuribe}, {Schneider} \&
  {Ferrara}}{{Omukai} et~al.}{2005}]{Omukai2005}
{Omukai} K.,  {Tsuribe} T.,  {Schneider} R.,    {Ferrara} A.,  2005, \apj, 626,
  627

\bibitem[\protect\citeauthoryear{Page, Brin, Motwani \& Winograd}{Page
  et~al.}{1999}]{Page1999}
Page L.,  Brin S.,  Motwani R.,    Winograd T.,  1999, Technical Report
  1999-66, The PageRank Citation Ranking: Bringing Order to the Web.,
  \verb+http://ilpubs.stanford.edu:8090/422/+.
Stanford InfoLab

\bibitem[\protect\citeauthoryear{{Press}, {Teukolsky}, {Vetterling} \&
  {Flannery}}{{Press} et~al.}{1992}]{NumericalRecipes}
{Press} W.~H.,  {Teukolsky} S.~A.,  {Vetterling} W.~T.,    {Flannery} B.~P.,
  1992, {Numerical recipes in FORTRAN. The art of scientific computing}.
Cambridge University Press

\bibitem[\protect\citeauthoryear{{Ruffle}, {Rae}, {Pilling}, {Hartquist} \&
  {Herbst}}{{Ruffle} et~al.}{2002}]{Ruffle02}
{Ruffle} D.~P.,  {Rae} J.~G.~L.,  {Pilling} M.~J.,  {Hartquist} T.~W.,
  {Herbst} E.,  2002, \aap, 381, L13

\bibitem[\protect\citeauthoryear{{Schultz}}{{Schultz}}{2006}]{SchultzPhdThesis}
{Schultz} M., , 2006, {``Dimension Reduction of Biochemical Network Models:
  Implementation and Comparison of Different Algorithms''},
  \url{http://www2.hu-berlin.de/biologie/theorybp/docs/bsc\_marvin\_schulz.pdf}

\bibitem[\protect\citeauthoryear{{Te{\c s}ileanu}, {Mignone} \&
  {Massaglia}}{{Te{\c s}ileanu} et~al.}{2008}]{Tesileanu2008}
{Te{\c s}ileanu} O.,  {Mignone} A.,    {Massaglia} S.,  2008, \aap, 488, 429

\bibitem[\protect\citeauthoryear{Tupper}{Tupper}{2002}]{Tupper2002}
Tupper P.~F.,  2002, Bit Numerical Mathematics, 42, 447

\bibitem[\protect\citeauthoryear{Vajda, Valko \& Turányi}{Vajda
  et~al.}{1985}]{Vajda1985}
Vajda S.,  Valko P.,    Turányi T.,  1985, International Journal of Chemical
  Kinetics, 17, 55

\bibitem[\protect\citeauthoryear{{Wakelam} \& {Herbst}}{{Wakelam} \&
  {Herbst}}{2008}]{Wakelam2008}
{Wakelam} V.,  {Herbst} E.,  2008, \apj, 680, 371

\bibitem[\protect\citeauthoryear{{Wiebe}, {Semenov} \& {Henning}}{{Wiebe}
  et~al.}{2003}]{Wiebe2003}
{Wiebe} D.,  {Semenov} D.,    {Henning} T.,  2003, \aap, 399, 197

\bibitem[\protect\citeauthoryear{{Yamasawa}, {Habe}, {Kozasa}, {Nozawa},
  {Hirashita}, {Umeda} \& {Nomoto}}{{Yamasawa} et~al.}{2011}]{Yamasawa2011}
{Yamasawa} D.,  {Habe} A.,  {Kozasa} T.,  {Nozawa} T.,  {Hirashita} H.,
  {Umeda} H.,    {Nomoto} K.,  2011, \apj, 735, 44

\end{thebibliography}


\bsp

\label{lastpage}

\end{document}